\documentclass[aps, prd, preprintnumbers, tightenlines, showpacs,nofootinbib,superscriptaddress, 11pt]{revtex4}

\usepackage{amssymb}
\usepackage{amsmath}
\usepackage{epsfig}

\begin{document}

\title{Sudakov Resummations in Mueller-Navelet Dijet Production}

\author{A. H. Mueller}
\affiliation{Department of Physics, Columbia University, New York, NY 10027, USA}

\author{Lech Szymanowski}
\affiliation{National Centre for Nuclear Research (NCBJ), Warsaw, Poland}

\author{Samuel Wallon}
\affiliation{Laboratoire de Physique Th\'{e}orique, UMR 8627, CNRS, Univ. Paris Sud, Universit\'{e} Paris-Saclay, 91405 Orsay, France}
\affiliation{UPMC Univ. Paris 06, Facult\'e de Physique, 4 place Jussieu, 75252 Paris Cedex 05, France}

\author{Bo-Wen Xiao}
\affiliation{Key Laboratory of Quark and Lepton Physics (MOE) and Institute
of Particle Physics, Central China Normal University, Wuhan 430079, China}

\author{Feng Yuan}
\affiliation{Nuclear Science Division, Lawrence Berkeley National
Laboratory, Berkeley, CA 94720, USA}

\begin{abstract}
In high energy hadron-hadron collisions, dijet production with large rapidity separation proposed by Mueller and Navelet, is one of the most interesting processes which can help us to directly access the well-known Balitsky-Fadin-Kuraev-Lipatov evolution dynamics. The objective of this work is to study the Sudakov resummation of Mueller-Navelet jets. Through the one-loop calculation, Sudakov type logarithms are obtained for this process when the produced dijets are almost back-to-back. These results could play an important role in the phenomenological study of dijet correlations with large rapidity separation at the LHC. 
\end{abstract}
\preprint{LPT-Orsay-15-85}
\maketitle

\section{Introduction}
In high energy collisions, small-$x$ evolution provides the QCD description of the dynamics of gluon evolution in the high energy limit when the longitudinal momentum fraction $x$ of partons is small. Due to the enhancement of the Bremsstrahlung radiation of small-$x$ gluons, high energy scattering amplitudes are expected to rise rapidly as collision energy increases. The rise of the resulting scattering cross sections can also be seen from the solution of Balitsky-Fadin-Kuraev-Lipatov (BFKL) evolution equation\cite{Balitsky:1978ic} which increases as the rapidity interval $Y=\ln1/x$ increases. The important feature of BFKL evolution is that the resulting cross section grows as $e^{(\alpha_P-1)Y}$, with $\alpha_P-1=\frac{4\alpha_sN_c}{\pi}\ln2$ at leading order. This behaviour essentially is equivalent to the exchange of a pomeron, thus sometimes the rise of gluon density and cross sections is attributed to the so-called BFKL pomeron.

\begin{figure}[tbp]
\begin{center}
\includegraphics[height=6.7cm]{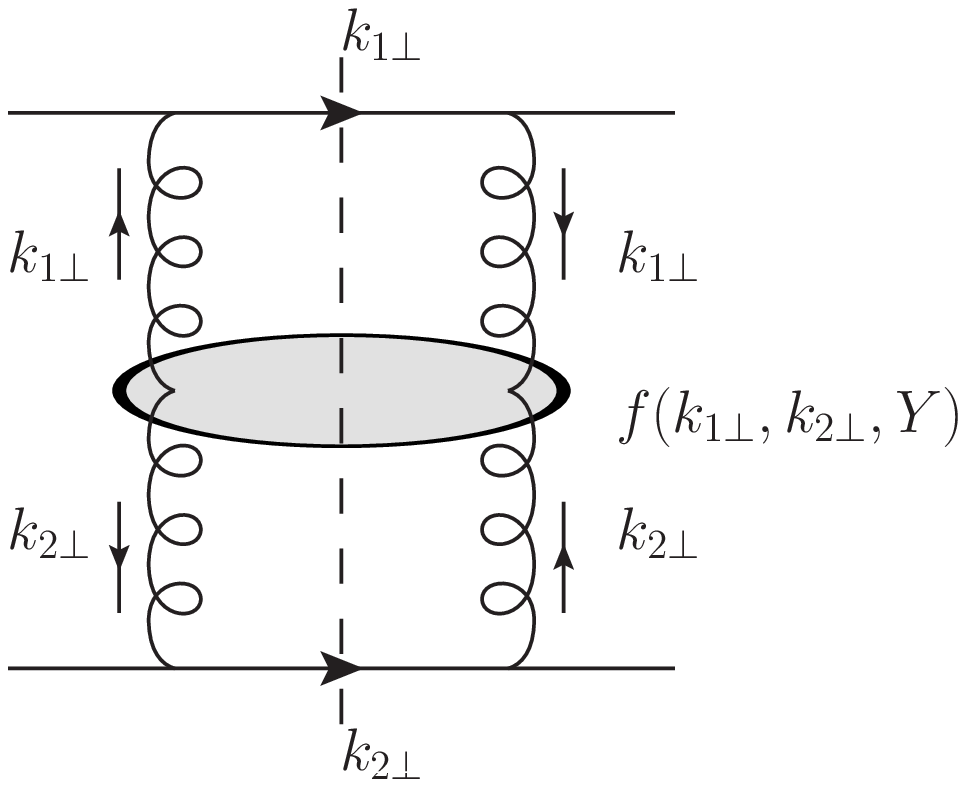}\includegraphics[height=6.3cm]{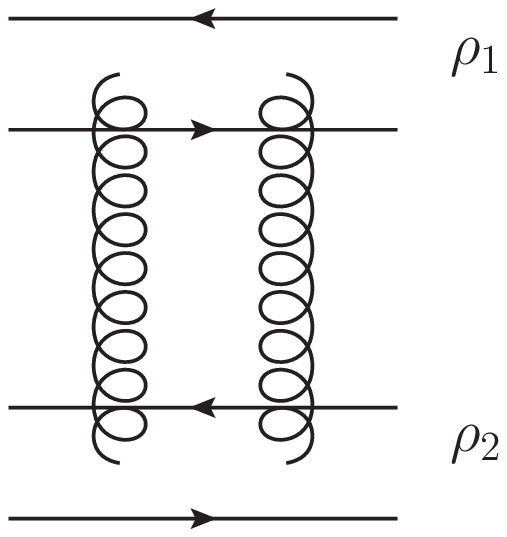}
\end{center}
\caption[*]{Left figure: Illustration of the quark-quark channel Mueller-Navelet dijet production in momentum space. Right figure: same process in the dipole model.}
\label{di}
\end{figure}

In high energy proton-proton collisions, inclusive dijets productions with large rapidity separation
\begin{equation}
p+p \to \left.\textrm{jet}_1 (y_1, k_{1\perp})\right|_{y_1>0}+ \left.\textrm{jet}_2(y_2, k_{2\perp})\right|_{y_2<0} +X, 
\end{equation}
which is known as Mueller-Navelet jets production, are particularly interesting for studying the properties of the BFKL pomeron and small-$x$ gluon evolution in the era of the LHC. Here $y_i$ and $k_{i\perp}$ represent the rapidities and transverse momenta of the produced jets. At the leading order (LO), the differential cross section of this process\cite{Mueller:1986ey} can be written as 
\begin{equation}
\frac{d\sigma}{dy_1 dy_2 d^2k_{1\perp}d^2k_{2\perp}} = \left[x_1g(x_1)+\frac{4}{9} x_1q(x_1) \right]\left[x_2g(x_2)+\frac{4}{9} x_2q(x_2) \right] \sigma_0 (k_{1\perp}, k_{2\perp}) f(k_{1\perp}, k_{2\perp}, Y), \label{mnlo}
\end{equation}
where $\sigma_0 (k_{1\perp}, k_{2\perp}) =\left(\frac{\alpha_sC_A}{\pi}\right)^2\frac{\pi}{2 k_{1\perp}^2k_{2\perp}^2}$, and $f(k_{1\perp}, k_{2\perp}, Y)$ obeys the momentum space representation of the BFKL evolution equation with rapidity interval $Y=y_1-y_2$. The physical picture of the LO Mueller-Navelet jets production is as follows: one parton with longitudinal momentum fraction $x_1=\frac{k_{1\perp}}{\sqrt{s}}e^{y_1}$ from the projectile proton with positive rapidity and another parton with longitudinal momentum fraction $x_2=\frac{k_{2\perp}}{\sqrt{s}}e^{-y_2}$ from the target proton with negative rapidity exchange a BFKL pomeron, which is characterized by the so-called BFKL pomeron propagator $f(k_{1\perp}, k_{2\perp}, Y)$, and eventually becomes two jets at rapidity $y_1$ and $y_2$, respectively. This is illustrated as in the left figure of Fig.~\ref{di}. We suppose that the rapidity interval $Y=y_1-y_2$ is so large that $x_1$ and $x_2$ are reasonably large. Therefore, the use of the collinear parton distributions, which neglect the transverse momenta of partons inside protons, can be justified. In this scenario, the transverse momentum imbalance of these two jets is due to the small-$x$ gluon radiation which is resummed by the BFKL evolution equation, and the azimuthal angular correlation is solely determined by the BFKL dynamics, namely $f(k_{1\perp}, k_{2\perp}, Y)$. 

Eq.~(\ref{mnlo}) gives the dominant contribution when $Y$ is sufficiently large. For not so large $Y$, if we neglect the parton shower, namely the Sudakov effects, we expect that these two jets are almost back-to-back in the azimuthal plane due to hard scattering. If we roughly fix the transverse momenta of the jets and increase their rapidity interval $Y$, we then have more and more gluons radiated with randomized transverse momenta due to the increment of the BFKL evolution. Thus, these two jets get less and less correlated, and may even become completely decorrelated at asymptotically large $Y$. Recently, the CMS collaboration at the LHC
has measured the dijet azimuthal correlation with large rapidity separation 
between the jets, which has been interpreted as the BFKL evolution (resummation) effects. This pattern of decorrelation with increasing $Y$ has been qualitatively observed by the CMS collaboration\cite{CMS:2013eda} at the LHC.

One should however note that this pattern can be significantly modified when including corrections to the impact factors describing the production of the two jets. In order to quantitatively compare with data for Mueller-Navelet jets, one needs to compute the one-loop diagrams and also include the next-to-leading order(NLO) contributions, besides the correction from the NLO BFKL evolution\cite{Fadin:1998py}. This has been intensively studied in the last two decades by several groups\cite{Ciafaloni:1998kx,Ciafaloni:1998hu, Bartels:2001ge, Bartels:2002yj, Caporale:2011cc, Colferai:2010wu, Caporale:2013uva}. Reasonably good agreement between the NLO calculation and the CMS data has been achieved\cite{Ducloue:2013hia, Ducloue:2013bva, Caporale:2014gpa, Celiberto:2015yba}. 

In light of recent development\cite{Mueller:2012uf, Mueller:2013wwa, Balitsky:2015qba, Marzani:2015oyb} of Sudakov resummation in small-$x$ formalism, by reexamining the one-loop diagrams associated with this process, we find that there also exist Sudakov type logarithms in Mueller-Navelet jets production in the configuration in which the produced jets are almost back-to-back. It was found that the resummation of Sudakov type logarithms and small-$x$ logarithm can be performed separately when two scales are present. (In this particular process, we have the average transverse momentum $P_\perp\simeq |k_{1\perp}|\simeq |k_{2\perp}|$ which characterizes the hard scattering and the dijet momentum imbalance $\vec{q}_\perp\equiv \vec{k}_{1\perp}+\vec{k}_{2\perp}$ which is due to gluon radiation. In the back-to-back configuration, it is clear that $P_\perp^2 \gg q_\perp^2$, which generates the Sudakov type logarithms, such as $\alpha_s \ln^2 \frac{P_\perp^2}{q_\perp^2}$.) We shall use the same technique  developed in Ref.~\cite{Mueller:2013wwa} in the following calculation to derive the Sudakov double logarithms for Mueller-Navelet jets production in $pp$ collisions in the Coulomb gauge which treats both the projectile and target protons symmetrically. 

We expect that Sudakov resummation introduces the suppression of back-to-back configurations, which is playing a similar role as the BFKL evolution in terms of dijet decorrelation. Of course, at asymptotically high energy with extremely large rapidity separation $Y$, the BFKL part is dominant and Sudakov suppression is presumed to be negligible. Nevertheless, at present LHC energy and kinematical regime where the measurement is made, we believe that these two effects should be taken into account together in order to achieve a better description of the LHC data. 

The original derivation of BFKL evolution was achieved in momentum space, which motivates the idea of $k_t$ factorization in high energy scatterings. Later, the color dipole picture of the BFKL pomeron in coordinate space was found in Ref.~\cite{Mueller:1993rr, Chen:1995pa}, and the exact equivalence between the color dipole model and the original BFKL results was verified afterwards\cite{Navelet:1997tx}. Since it is mostly convenient to perform Sudakov resummation in coordinate space in order to take the momentum conservation of arbitrary number of gluons into account, the color dipole model is then the natural choice of framework to work with. As illustrated in the right figure of Fig.~\ref{di}, by using Fourier transform with proper normalization, we can convert the above expression in Eq.~(\ref{mnlo}) into the so-called $T$-matrix, which obeys the coordinate space representation of the BFKL equation in the color dipole model. Therefore, in the following discussion, we will derive the Sudakov double logarithm from the one-loop calculation of the Mueller-Navelet dijet production by using the color dipole model, and discuss the resummation of Sudakov logarithms. 

The rest of the paper is organized as follows. In Sec. II, we briefly discuss the lowest order dipole-dipole scattering amplitude, which helps us to fix the normalization with the momentum space expression. Then, we use the dipole splitting function to compute one-loop diagrams in coordinate space and derive the Sudakov double logarithm in Sec. III. In Sec. IV, we give an intuitive discussion on the origin of this Sudakov factor and discuss its implications. At last, we conclude in Sec. V, and provide some discussion on the emergence of the Sudakov factor from the collinear factorization point of view in the Appendix.

\section{Leading order cross section in the color dipole model}
In this section, we would like to specify our normalization and compare the dipole model approach with the usual BFKL approach in momentum space and collinear factorization results. First of all, let us compute the LO dipole-dipole scattering amplitude and show that it is equivalent to the momentum space results. Throughout the paper, we work in light-cone coordinates and define light cone variables as $p^+=\frac{p^0+p^z}{\sqrt{2}}$ and $p^-=\frac{p^0-p^z}{\sqrt{2}}$.
As shown in Fig.~\ref{di}, the leading order cross section can be written down as follows
\begin{equation}
\frac{d\sigma(qq\to qq)}{dy_1 dy_2 d^2k_{1\perp}d^2k_{2\perp}}=x_1 f_1(x_1) x_2 f_2(x_2) \int \frac{d^2\rho_1 d^2 \rho_2}{(2\pi)^4}e^{-ik_{1\perp}\cdot \rho_1-ik_{2\perp}\cdot \rho_2}
T(\rho_1,\rho_2,Y)\ , \label{lo}
\end{equation}
where $T(\rho_1,\rho_2,Y)$ represents the scattering matrix between two coordinate dipoles with size $\rho_1$ and $\rho_2$ as depicted in Fig.~\ref{di}. In the center of mass frame, $x_1=\frac{k_{1\perp}}{\sqrt{2}p^+_1}e^{y_1}$, $x_2=\frac{k_{2\perp}}{\sqrt{2}p^-_2}e^{-y_2}$, $Y=y_1-y_2=\ln\frac{x_1 x_2 S}{\tilde P_\perp^2}$ with $\tilde P_\perp^2\equiv |k_{1\perp}| | k_{2\perp}|$ and center of mass energy $S=2 p^+_1 p_2^-$. For the quark-quark channel, we just need to set $f_1=q_1$ and $f_2=q_2$. For other channels, we just need to use the corresponding parton distributions and put proper color factor for the $T-$matrix. Therefore, let us focus on the one-loop calculation for the quark-quark channel, since the derivation for other channels is rather similar.   
At the lowest order without any BFKL evolution, namely without any gluon radiation, one finds that it is given by the dipole-dipole scattering cross section which can be written as \cite{Mueller:2001fv}
\begin{equation}
T_0(\rho_1,\rho_2) = \alpha_s^2\frac{N_c^2-1}{4N_c^2}\int\frac{d^2 l_\perp}{l_\perp^4}\left(2-e^{-il_\perp\cdot \rho_1}-e^{il_\perp\cdot \rho_1}\right)\left(2-e^{-il_\perp\cdot \rho_2}-e^{il_\perp\cdot \rho_2}\right)
\ ,
\end{equation}
which gives 
\begin{equation}
\frac{d\sigma(qq\to qq)}{dy_1 dy_2 d^2k_{1\perp}d^2k_{2\perp}}=x_1 f_1(x_1) x_2 f_2(x_2) \frac{N_c^2-1}{N_c^2} \frac{\alpha_s^2}{k_{1\perp}^2k_{2\perp}^2}\delta^{(2)}(k_{1\perp}+k_{2\perp})\ . \label{loc}
\end{equation}
The above results is equivalent to Eq.~(\ref{mnlo}) for the quark-quark channel once we set $f(k_{1\perp}, k_{2\perp}, Y)=\delta^{(2)}(k_{1\perp}+k_{2\perp})$ when $Y=0$. Moreover, we can obtain exactly the same results as in lowest order collinear factorization calculation after taking into account 
\begin{equation}
\frac{d\sigma(qq\to qq)}{\pi dt}=\frac{N_c^2-1}{4N_c^2}\frac{2\alpha_s^2}{s^2}\left[\frac{s^2+u^2}{t^2}+\frac{t^2+u^2}{s^2}-\frac{2}{3}\frac{u^2}{st}\right]\ \simeq \frac{N_c^2-1}{N_c^2}\frac{\alpha_s^2}{t^2}
\end{equation}
in the high energy $-t \ll s \simeq -u$ limit, where $-t=k_{1\perp}^2=k_{2\perp}^2$ and $s=x_1 x_2 S$. In addition, it has been shown that the dipole model is completely equivalent to the BFKL Green's function approach. By relating the BFKL pomeron propagator $f(k_{1\perp}, k_{2\perp}, Y)$ to the Fourier transform of $T(\rho_1,\rho_2,Y)$, we can easily demonstrate that Eq.~(\ref{lo}) is equivalent to the LO formula with BFKL evolution in Ref.~\cite{Mueller:1986ey} (See also e.g., Refs.~\cite{Colferai:2010wu, Caporale:2011cc}). Since the systematic resummation of Sudakov double logarithms can conveniently be done in coordinate space, we choose to do the calculation in the dipole model. 

\section{One-loop calculation and the Sudakov double logarithms}
At one-loop level, on top of the LO diagram, we need to consider the radiation of an extra gluon. In principle, this extra gluon radiation can occur anywhere in Fig.~\ref{di}. To perform the dipole model calculation at one-loop, we choose to employ the Coulomb gauge following Ref.~\cite{Jaroszewicz:1980mq} which allows us to simplify the one-loop calculation in leading logarithm approximation (LLA), by removing all the diagrams with additional gluon exchanges between two partons with large rapidity intervals, which are suppressed in high energy limit. That is to say, at the level of LLA in high energy scatterings, the dominant contributions always come from the diagrams with two vertical gluon exchanges between the projectile and target protons.  

Let us suppose the four-momentum of the radiated gluon is $(l^+, l^-, l_\perp)$. As long as $l^+ >l^-$, namely $l^+ >\frac{l_\perp}{\sqrt{2}}$ for real gluons, the Coulomb gauge is equivalent to the light cone gauge with $A^+=0$\cite{Mueller:2001fv}, which allows us to compute gluon radiation from the right moving quark in the dipole model. As to the region $l^+  <l^-$ in the phase space of the radiated gluon, the Coulomb gauge reduces to the light cone gauge with $A^-=0$, which indicates that the gluon radiation is originated from the left moving quark. These two regions are completely symmetric, thus we can just compute the former and multiply by a factor of $2$ to take into account the latter. With this choice of gauge, we still can use the dipole splitting kernel which is derived in the light cone gauge with the above corresponding constraints.

\subsection{The Derivation of the Sudakov Factor}
\begin{figure}[tbp]
\begin{center}
\includegraphics[width=6.2cm]{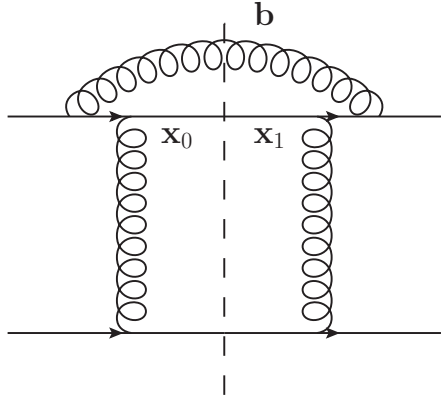}
\end{center}
\caption[*]{Real diagram with the initial state gluon radiation.}
\label{real1}
\end{figure}
At one-loop order, let us first consider the diagram as shown in Fig.~\ref{real1} in the eikonal approximation, which gives 
\begin{equation}
x_2 f_2(x_2) \frac{\alpha_s C_F}{(2\pi)^2}\int_{x_1}^1 d\xi x f_1(x)  \int \frac{d^2\textbf x_{10} d^2\textbf b d^2 \rho_2}{(2\pi)^6}e^{-ik_{1\perp}\cdot \textbf x_{10}-ik_{2\perp}\cdot \rho_2}
 \sum_{\lambda\alpha\beta}
\psi^{\lambda\ast}_{\alpha\beta}(\textbf x_0-\textbf b)\psi^\lambda_{\alpha%
\beta}(\textbf x_1-\textbf b)T(\textbf x_{10},\rho_2,Y) \ , \label{re1}
\end{equation}
where the two-dimensional coordinates of active partons are labeled in Fig.~\ref{real1} and $\textbf x_{10}\equiv \textbf x_1-\textbf x_0$. The longitudinal momentum fraction of the incoming quark $x$ is no longer fixed, instead, now it becomes $x_1/\xi$ with $\xi >x_1$. For the right-moving massless quark with no initial transverse momentum and initial longitudinal momentum $p^+$, 
the splitting wave function of $q\to q+g$ in transverse coordinate space can be cast into (see e.g. Ref.~\cite{Marquet:2007vb})
\begin{equation}
\psi^\lambda_{\alpha\beta}(p^+,k^+,u_{\perp})=2\pi i\sqrt{\frac{2}{k^+}}%
\begin{cases}
\frac{u_{\perp}\cdot\epsilon^{(1)}_\perp}{u_{\perp}^2}(\delta_{\alpha-}%
\delta_{\beta-}+\xi\delta_{\alpha+}\delta_{\beta+}), & \lambda=1, \\
\frac{u_{\perp}\cdot\epsilon^{(2)}_\perp}{u_{\perp}^2}(\delta_{\alpha+}%
\delta_{\beta+}+\xi\delta_{\alpha-}\delta_{\beta-}), & \lambda=2,%
\end{cases}
\ ,  \label{wfun}
\end{equation}
where $\lambda$ represents the gluon polarization, $\alpha,\beta$ indicate
helicities for the incoming and outgoing quarks, and $1-\xi=\frac{k^+}{p^+}
$ is defined as the longitudinal momentum fraction of the incoming quark carried by the radiated gluon. Here $u_\perp$ is the transverse separation of the quark-gluon pair, and it is conjugate to their relative momentum. 
When $\xi\to 1$, the radiated gluon becomes very soft. By performing the Fourier transform, it is straightforward to show that Eq.~(\ref{re1}) can be converted to
\begin{eqnarray}
x_2 f_2(x_2) \frac{\alpha_s C_F}{2\pi^2}\int_{x_1}^1 d\xi \frac{1+\xi^2}{1-\xi}x f_1(x) \int\frac{d^2 l_\perp}{l_\perp^2} e^{-il_\perp\cdot \textbf x_{10}} \int \frac{d^2\textbf x_{10}  d^2 \rho_2 }{(2\pi)^4}e^{-ik_{1\perp}\cdot\textbf x_{10}-ik_{2\perp}\cdot \rho_2} T(\textbf x_{10},\rho_2,Y). \label{re11}
\end{eqnarray}
Now our task is to evaluate Eq.~(\ref{re11}). First of all, according to the definition of the plus-function, one can write
\begin{equation}
\int_{x_1}^1 d\xi  \frac{1+\xi^2}{1-\xi}x f_1(x) =\int_{x_1}^1 d\xi  \frac{1+\xi^2}{(1-\xi)_+}x f_1(x)+x_1f_1(x_1)\int_{0}^1 d\xi \frac{2}{1-\xi}.
\end{equation}
As demonstrated before in Ref.~\cite{Mueller:2013wwa}, the first term in the above equation corresponds to the renormalization of the collinear parton distribution, since it only contains collinear singularities after being put back to Eq.~(\ref{re11}). (The finite part can be put into the NLO hard factor.)
As to the second term, after taking into account the constraint $1-\xi=\frac{l^+}{x_1p_1^+}>\frac{l_\perp}{\sqrt{2}x_1p_1^+}$ due to the use of Coulomb gauge with respect to the above $\xi$ integration, we can obtain
\begin{equation}
\int d\xi \frac{2}{1-\xi}=2\ln\frac{\sqrt{2}x_1p_1^+}{l_\perp}=2\ln\frac{\sqrt{2}x_1p_1^+}{\tilde{P}_\perp}+\ln\frac{\tilde{P}_\perp^2}{l_\perp^2}.
\end{equation}
Similarly, one can consider the gluon radiation from the left moving quark quark with the large $-$ momentum component $x_2 p_2^-$ and obtain
\begin{equation}
2\ln\frac{\sqrt{2}x_2p_2^-}{\tilde{P}_\perp}+\ln\frac{\tilde{P}_\perp^2}{l_\perp^2}.
\end{equation}
Adding these two contributions together, we find that the first part gives contribution which is proportional to $\alpha_s\ln\frac{x_1 x_2 S}{\tilde P_\perp^2}=\alpha_s Y$. It is obvious that this corresponds to the BFKL evolution of the dipole-dipole scattering cross section. Following the same strategy\cite{Mueller:2013wwa}, one can demonstrate that the BFKL evolution equation can be derived after taking all the graphs into account. 

After renaming $\textbf x_{10}$ to $\rho_1$, the second part can be cast into
\begin{equation}
x_1 f_1(x_1) x_2 f_2(x_2) 4\alpha_s C_F \int\frac{d^2 l_\perp}{(2\pi)^2l_\perp^2} \ln\frac{\tilde{P}_\perp^2}{l_\perp^2} \int \frac{d^2 r_\perp d^2 R_\perp}{(2\pi)^4}e^{-i\left(q_{\perp}+l_\perp\right)\cdot R_\perp -iP_{\perp}\cdot r_\perp} T(\rho_1,\rho_2,Y)\ 
\end{equation}
where $q_\perp\equiv k_{1\perp}+k_{2\perp}$, $P_{\perp}\equiv \frac{1}{2}(k_{1\perp}-k_{2\perp})$, $r_\perp\equiv \rho_1 -\rho_2$ and $R_{\perp}\equiv \frac{1}{2}(\rho_1+\rho_2)$. In the above equation, we have neglected $l_\perp$ (which is the order of $q_\perp$) as compared to $P_\perp$. The change of variables here is important to our calculation, since $q_\perp$ and $P_\perp$ are the most convenient and relevant variables in the back-to-back limit. Furthermore, in the back-to-back dijet limit, since $P_\perp^2 \simeq \tilde P_\perp^2 \gg q_\perp$, we do not distinguish between $P_\perp$ and $\tilde P_\perp$.
In the leading power approximation, we neglect all contributions which are of order of $q_\perp^2/P_\perp^2$. This allows us to also neglect $l_\perp$ as compared to $P_\perp$. When $l_\perp$ is as large as $P_\perp$, one can easily see that the resulting contribution is then power suppressed. 

Now the integral in question is
\begin{equation}
 \int\frac{d^2 l_\perp}{(2\pi)^2l_\perp^2} e^{-il_{\perp
}\cdot R_\perp}\ln\frac{\tilde{P}_\perp^2}{l_\perp^2}\Rightarrow \mu^{2\epsilon}\int\frac{\text{d}^{2-2\epsilon}l_{\perp}}{(2\pi)^{2-2\epsilon}}e^{-il_{\perp
}\cdot R_\perp}\frac{1}{l_{\perp}^{ 2}}\ln\frac{\tilde{P}_\perp^2}{l_\perp^2}, \label{realsudakov}
\end{equation}
where we have changed the dimension of the integral from $2$ to $2-2\epsilon$ in order to isolate the expected soft-collinear divergence. In the $\overline{\textrm{MS}}$ scheme, we find that the above integral yields (see the appendix of Ref.~\cite{Mueller:2013wwa})
\begin{equation}
\frac{1}{4\pi}\left[\frac{1}{\epsilon^2}-\frac{1}{\epsilon}\ln\frac{\tilde{P}_\perp^2}{\mu^2}
+\frac{1}{2}\left(\ln\frac{\tilde{P}_\perp^2}{\mu^2}\right)^2-\frac{1}{2}\left(\ln\frac{\tilde{P}_\perp^2 R_\perp^2}{c_0^2}\right)^2 
-\frac{\pi^2}{12}\right] \ , \label{realc}
\end{equation}
where $c_0=2e^{-\gamma_E}$ and $\gamma_E$ is the Euler constant.

\begin{figure}[tbp]
\begin{center}
\includegraphics[width=16.0cm]{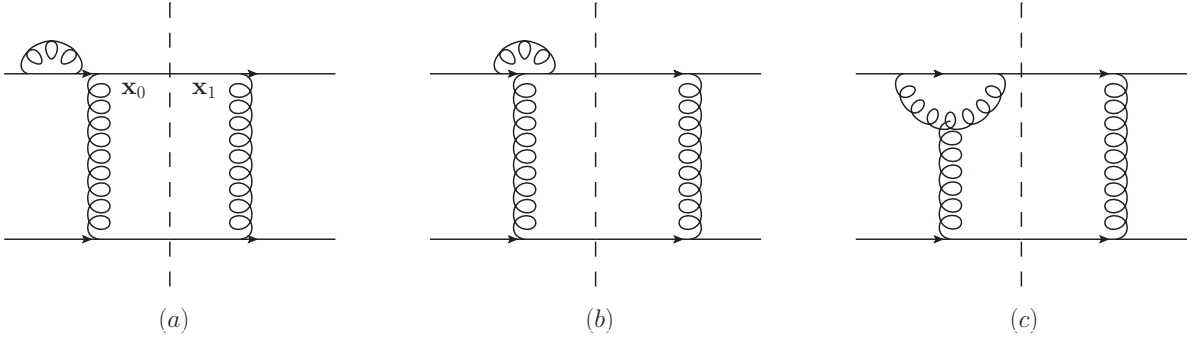}
\end{center}
\caption[*]{Virtual diagrams.}
\label{vir1}
\end{figure}

Now let us consider the virtual graphs as shown in Fig.~\ref{vir1}. In principle, we need to take all these three graphs in Fig.~\ref{vir1} into account. We shall not present the complete calculation here, since it is a bit tedious.\footnote{In the case of Higgs productions and dijet productions in $pA$ collisions, detailed computations of virtual diagrams are presented in Ref.~\cite{Mueller:2013wwa}. The technique needed to perform such complete calculation for Mueller-Navelet jets is akin to that used in Higgs productions and dijet productions.}  We found a quick way to obtain the total virtual contribution by using the simple fact that the ultra-violet divergence should cancel between these three graphs. Due to this cancellation, it is natural to just simply assume that these three virtual graphs completely cancel in the ultra-violet region where $l_\perp > \tilde{P}_\perp$. On the other hand, in the $l_\perp <\tilde{P}_\perp$ region, we find that graph (b) and (c) are power suppressed, thus can be neglected. Therefore, the only important contribution comes from graph (a) with the constraint $l_\perp <\tilde{P}_\perp$, which amazingly gives the identical result as the complete evaluation of all three graphs. 

Again, using the dipole model and the Fourier transform, it is straightforward to find that graph (a) in Fig.~\ref{vir1} gives the following contribution
\begin{equation}
\int_0^1 d\xi \frac{1+\xi^2}{1-\xi}\int\frac{d^2 l_\perp}{(2\pi)^2l_\perp^2}=\int_0^1 d\xi \frac{1+\xi^2}{(1-\xi)_+}\int\frac{d^2 l_\perp}{(2\pi)^2l_\perp^2}+\int\frac{d^2 l_\perp}{(2\pi)^2l_\perp^2} \int_0^1 d\xi \frac{2}{1-\xi}.
\end{equation}
Similarly, taking the Coulomb gauge constraint into account, the second term gives 
\begin{equation}
\int_0^{1-\frac{l_\perp}{\sqrt{2}x_1p_1^+}}  \frac{2d\xi }{1-\xi}=2\ln\frac{\sqrt{2} x_1p_1^+}{l_\perp}=2\ln\frac{\sqrt{2}x_1p_1^+}{\tilde{P}_\perp}+\ln\frac{\tilde{P}_\perp^2}{l_\perp^2}.
\end{equation}
Following the same procedure, we identify the first part as the contribution to the BFKL evolution, and the second part as the contribution to the Sudakov factor, which can be cast into
\begin{equation}
\left. \int\frac{d^2 l_\perp}{(2\pi)^2l_\perp^2}\ln\frac{\tilde{P}_\perp^2}{l_\perp^2}\right|_{l_\perp<\tilde{P}_\perp}=\frac{1}{4\pi}\left[\frac{1}{\epsilon^2}-\frac{1}{\epsilon}\ln\frac{\tilde{P}_\perp^2}{\mu^2}
+\frac{1}{2}\left(\ln\frac{\tilde{P}_\perp^2}{\mu^2}\right)^2
-\frac{\pi^2}{12}\right] .\label{virc}
\end{equation}
As commented above, we have made an ultra-violet cut $l_\perp <\tilde{P}_\perp$ in the above integration by using the knowledge that the large $l_\perp$ region will be cancelled by other virtual diagrams.

By adding the real and virtual contributions together, one can easily find that the soft and collinear divergences cancel, and the remaining Sudakov factor is\footnote{We always find this imperfect cancellation between real and virtual contributions, as long as we require that back-to-back dijets with fixed momenta $k_{1\perp}$ and $k_{2\perp}$ are produced. Qualitatively speaking,  Sudakov double logarithms always arise due to the incomplete cancellation of the soft-collinear region of phase space between real and virtual graphs. As far as the back-to-back dijet correlation is concerned, such incomplete cancellation is bound to occur, since a certain constraint should be put on the real graphs to generate the desired dijet configuration, while virtual graphs have no constraint at all. On the other hand, if one integrates over the full phase space of one of the dijet momenta (e.g. $k_{2\perp}$) at one-loop level, one should find that Sudakov double logarithms are absent. This implies that generic dijet productions at one-loop level can not be viewed as productions of two independent jets.}
\begin{equation}
S_{qq \to qq}=-\frac{\alpha_s C_F}{2\pi}\ln^2\frac{\tilde{P}_\perp^2 R_\perp^2}{c_0^2},
\end{equation}
which becomes the so-called Sudakov suppression factor after exponentiation due to multiple gluon radiation. Since one needs to impose a delta function due to the conservation of transverse momentum when performing the resummation of arbitrary number of Sudakov gluon radiations, it is common practice to do the Fourier transform of that delta function and find that the Sudakov factor naturally exponentiates in the coordinate space. It is also interesting to note that this result with the effective colour factor $C_F$ agrees with the empirical formula\cite{Mueller:2013wwa} for the Sudakov double logarithmic factor which implies that each incoming quark contributes $\frac{1}{2}C_F$ to the effective colour factor. At the end of the day, in the back-to-back limit, we find the cross section of Mueller-Navelet jets production 
\begin{equation}
\frac{d\sigma(qq\to qq)}{dy_1 dy_2 d^2q_{\perp}d^2P_{\perp}}=x_1 f_1(x_1) x_2 f_2(x_2) \int \frac{d^2 r_\perp d^2 R_\perp}{(2\pi)^4}e^{-iq_{\perp}\cdot R_\perp -iP_{\perp}\cdot r_\perp} e^ {-\frac{\alpha_s C_F}{2\pi} \ln^2 \frac{P_\perp^2 R_\perp^2}{c_0^2}}
T(\rho_1,\rho_2,Y)\ , \label{re}
\end{equation}
where naturally a convolution of the BFKL evolved $T$-matrix together with the Sudakov factor in coordinate space occurs. We find that both the Sudakov resummation and BFKL evolution suppress the back-to-back configuration of dijet productions. On the other hand, we expect that the Sudakov factor is important when the rapidity separation $Y$ is not too large, while the BFKL pomeron exchange dominates when $Y$ is asymptotically large. At the present LHC kinematics, we believe that both effects should be taken into account. 

In addition, simply taking the difference in color factors into account, it is straightforward to generalize the above calculation and compute the Sudakov double logarithms for other dominant channels such as $qg \to qg$ and $gg \to gg$ as follows
\begin{eqnarray}
&& S_{qg \to qg}=-\frac{\alpha_s (C_F+C_A)}{4\pi}\ln^2\frac{\tilde{P}_\perp^2 R_\perp^2}{c_0^2},\\
&& S_{gg \to gg}=-\frac{\alpha_s C_A}{2\pi}\ln^2\frac{\tilde{P}_\perp^2 R_\perp^2}{c_0^2}.
\end{eqnarray}

\subsection{Comments On Other Graphs}
\begin{figure}[tbp]
\begin{center}
\includegraphics[width=5.5cm]{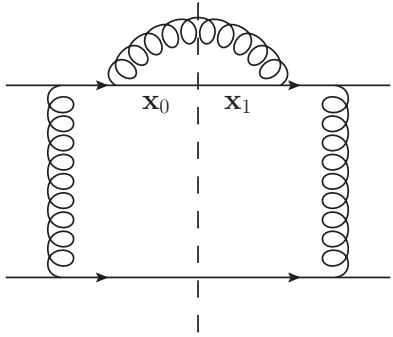}
\end{center}
\caption[*]{Real diagram with final state gluon radiation.}
\label{ref1}
\end{figure}

In the derivation of the Sudakov double logarithm, we find that only the above considered graphs contribute while the rest of  one-loop graphs do not. Some of the diagrams, which contain interactions between the radiated gluon and the t-channel exchanged vertical gluon, are simply power suppressed by factors of $\frac{q_\perp^2}{P_\perp^2}$, while other graphs do not contain Sudakov type double logarithms. 

There is one type of one-loop diagrams as shown in Fig.~\ref{ref1}, in which the final state gluon is radiated. In principle, this type of diagrams could contribute the Sudakov factor as well. However, in this particular Coulomb gauge that we choose, this diagram only contain the small-$x$ evolution and jet cone contributions so long as the azimuthal angular deviation from the jets being back-to-back, $\phi\sim \frac{q_\perp}{P_\perp}$, is less than the jet size, $\delta$. We are going to use the same trick employed in Ref.~\cite{Mueller:2013wwa} to study this graph as follows. In the soft gluon limit $\xi_g\equiv 1-\xi \ll 1$, the contribution from Fig.~\ref{ref1} after factorizing out the LO contribution can be cast into
\begin{equation}
\frac{4\alpha_s C_F}{(2\pi)^2}\int \frac{d^2 l_\perp}{(l_\perp-\xi_g k_{1\perp})^2}e^{-il_\perp \cdot \textbf x_{01}} \int^1_{\frac{l_\perp}{\sqrt{2}x_1p_1^+}} \frac{d\xi_g}{\xi_g}.
\end{equation}
Since we are only interested in the correlation between the produced dijet, we can average over the azimuthal angle of the leading jet, say the azimuthal angle of $k_{1\perp}$. Using the identity 
\begin{equation}
\frac{1}{2\pi} \int_0^{2\pi} d \theta \frac{1}{1+a \cos \theta} =\frac{1}{\sqrt{1-a^2}},  \quad \textrm{with} \quad a<1,
\end{equation}
one can cast the above integral into
\begin{equation}
4\alpha_s C_F  \int \frac{\textrm{d} ^2l_\perp}{(2\pi)^2} \frac{1}{|l_\perp^2-\xi_g^2 k_{1\perp}^2|} e^{-il_\perp \cdot x_{01}} \int_{\frac{l_\perp}{\sqrt{2}x_1p_1^+}}^1 \frac{\textrm{d} \xi_g}{\xi_g} .
\end{equation}
Obviously, the above integration has a collinear singularity at $\xi_g=\frac{l_\perp}{P_\perp}$ which is expected since this comes from the region where the radiated gluon is collinear to the quark. Let us simply regularize this collinear singularity by putting a cutoff $\delta$ in the $\xi$ integral which gives
\begin{equation}
\int_{l_\perp/\sqrt{2}x_1p_1^+}^{\frac{l_\perp}{k_{1\perp}} (1-\frac{1}{2}\delta)} \frac{\textrm{d} \xi_g}{\xi_g} \frac{1}{l_\perp^2-\xi_g^2 k_{1\perp}^2}-\int_{\frac{l_\perp}{k_{1\perp}} (1+\frac{1}{2}\delta)}^{1}\frac{\textrm{d} \xi_g}{\xi_g}  \frac{1}{l_\perp^2-\xi_g^2 k_{1\perp}^2}\simeq \frac{1}{l_\perp^2} \left[\ln\frac{\sqrt{2}x_1p^+_1}{k_{1\perp}}+\frac{1}{2}\ln\frac{1}{\delta^2}\right], \label{finaljet}
\end{equation}
where the azimuthal cone size $\delta$ should depend on the angular resolution of the jet measurement. The above results contain only two terms which correspond to two kinds of different physics, namely, the energy evolution and the jet cone definition. In principle, we should do a rigorous calculation with proper definition of cone size $R\equiv \sqrt{\Delta y^2 + \Delta \phi^2}$, where $\Delta y$ and $\Delta \phi$ are the rapidity and azimuthal angle size of the jets, respectively. We have done this complete calculation and found the same conclusion. In practice, we normally choose $\delta \sim R \sim 1$. 

By taking into account the similar diagram for gluon radiation originating from the quark in the left moving proton, we obtain $\ln\frac{\sqrt{2}x_1p^+_1}{k_{1\perp}}+\ln\frac{\sqrt{2}x_2p^-_2}{k_{2\perp}}=Y$, which corresponds to the BFKL evolution of the scattering dipole cross section. The second term clearly represents the jet cone singularity, which can be regularised easily by using more rigorous jet cone definition. As compared to the calculation in Ref.~\cite{Mueller:2013wwa}, which is computed in $A^+=0$ light-cone gauge, the Sudakov contribution is now absent in the Coulomb gauge calculation presented here. One can employ a jet function and rigorously compute the jet cross section. Nevertheless, this is independent of the calculation for the Sudakov double logarithms. \footnote{In Ref.~\cite{Mueller:2013wwa}, we studied the Sudakov factors in $pA$ collisions with the use of $A^+=0$ light-cone gauge, where we found that the final state radiation does contain Sudakov double logarithms. On the other hand, for the current Mueller-Navelet jets productions problem in $pp$ collisions, we have to choose the Coulomb gauge which treats both the projectile and target protons symmetrically, and we find no Sudakov double logarithmic contributions from graphs with final state gluon radiations. Although the conclusions with respect to the contribution of final state gluon radiation are different, we believe that there is no potential contradiction here due to different choice of gauges.}

Let us comment on other diagrams which have not been discussed above. For example, there are interference diagrams of initial and final state gluon radiation, and so as far as the Sudakov factor is concerned, those graphs do not contribute to the Sudakov double logarithms. In total, there are nine different types of real graphs and three different types of virtual graphs. However, the rapidity divergent part of all graphs contribute to the BFKL evolution of the dipole scattering amplitude. We have checked explicitly the combination of all the graphs naturally gives the total colour factor $\frac{N_c}{2}$ for the BFKL evolution equation in the dipole model as indicated below. At the lowest order, there is no $Y$ dependence in the dipole-dipole scattering cross section. At one-loop order, we find that the energy dependence can be absorbed into the redefinition of $T(\rho_1,\rho_2, Y)$ which gives the $Y$ dependence as follows
\begin{equation}
T(\rho_1,\rho_2, Y) \equiv T_0(\rho_1,\rho_2) +\frac{\alpha_s N_c Y}{2\pi^2} \int \frac{d^2 \textbf b \rho_1^2}{\textbf b^2 (\rho_1-\textbf b)^2} \left[-T_0(\rho_1,\rho_2)+T_0(\rho_1-\textbf b,\rho_2)+T_0(\textbf b,\rho_2)\right]. \label{bfklsub}
\end{equation}
This exactly agrees with the dipole model version of the BFKL evolution. Due to boost invariance, we can either put all the evolution in the projectile or put it in the target. This is justified since the solution of BFKL equation $T(\rho_1,\rho_2, Y)$ is symmetric between the interchange of $\rho_1$ and $\rho_2$. 

Last but not least, we would like to comment on the collinear singularities which also appear in the one-loop calculation in certain diagrams. We find that the collinear singularities associated with the initial state gluon radiation should be subtracted from the one-loop calculation and put into the redefinition of the corresponding incoming collinear parton distributions, which naturally yields the scale evolution of collinear parton distributions. On the other hand, through rigorous calculations with proper definition of jets, we find that the final state collinear singularities always cancel between real and virtual graphs, since jets are infrared safe observables.  

\section{Heuristic derivation}

Based on the calculation that we conducted above and techniques developed in Ref.~\cite{Mueller:2013wwa}, we provide below a heuristic derivation of Sudakov double logarithms, which is much simpler. Without getting into much technical detail, we use the general physical picture of Sudakov factors to illustrate how they arise from the one-loop calculation. In general, Sudakov effects occur when physical systems have two distinct scales besides the collision energy. In this problem, these two scales are the dijet momentum imbalance $q_\perp\equiv k_{1\perp}+k_{2\perp}$ and the jet transverse momentum $P_\perp\simeq |k_{1\perp}|\simeq |k_{2\perp}|$. In the back-to-back configuration, kinematics require $P_\perp \gg q_\perp$. The Sudakov factor helps to resum the large logarithms of their ratio, which start to appear at one-loop order.

At one-loop order, we have one extra gluon as compared to the LO diagram and we need to integrate over its phase space. Let us divide the phase space into three regions, namely the infrared region $\mu<l_\perp < q_\perp$ with the infrared cutoff $\mu$, the ultra-violet region $l_\perp > P_\perp$ and the region in between with $q_\perp <l_\perp <P_\perp$. Roughly speaking, after taking care of the collinear divergence associated with initial parton distributions, the infrared divergences should cancel between the real and virtual graphs. Furthermore, the ultra-violet divergences always cancel among real diagrams and virtual graphs separately. 

The back-to-back dijets are characterized by $q_\perp$ and $P_\perp$ with $q_\perp \ll P_\perp$. That is to say, the azimuthal angular deviation of the dijet system from $\pi$ should be less than $\phi \sim \frac{q_\perp}{P_\perp} \ll 1$. For a given dijet configuration in the back-to-back limit, the phase space of real gluon radiation is limited only to the infrared region defined above while the virtual graphs are unrestricted. Therefore, we can compute the probability of the back-to-back dijet directly. The contribution from the real diagram with initial state gluon radiation is 
\begin{equation}
\frac{\alpha_s C_F }{\pi^2}  \int_{\mu^2}^{q_\perp^2} \frac{d^2l_\perp}{l_\perp^2} \int^{P_\perp}_{l_\perp} \frac{dl^+}{l^+},
\end{equation}
and the virtual contribution is 
\begin{equation}
-\frac{\alpha_s C_F }{\pi^2} \int_{\mu^2}^{P_\perp^2} \frac{d^2l_\perp}{l_\perp^2} \int^{P_\perp}_{l_\perp} \frac{dl^+}{l^+},
\end{equation}
where we have truncated the $\int \frac{dl^+}{l^+}$ integration at $P_\perp$, since we should identify the contribution from the interval $[P_\perp, p^+]$ as the BFKL type contribution. The corresponding logarithm is $\ln\frac{s}{P_\perp^2}$ after taking the gluon splitting from the left-moving quark into account. From the same consideration, we should also multiply a factor $2$ to the above contributions. Therefore, the Sudakov contribution is 
\begin{equation}
-2\frac{\alpha_s C_F }{\pi^2} \int_{q_\perp^2}^{P_\perp^2} \frac{d^2l_\perp}{l_\perp^2} \int^{P_\perp}_{l_\perp} \frac{dl^+}{l^+}=-\frac{\alpha_s C_F }{2\pi} \ln^2\frac{P_\perp^2}{q_\perp^2},
\end{equation}
which agrees with the results obtained above after setting $R_\perp \sim 1/q_\perp$. Furthermore, we can also compute the above Sudakov double logarithm by only considering the dijet configuration with the angular deviation greater than $\phi$, which gives the probability of gluon radiation with transverse momentum $l_\perp > q_\perp$, 
\begin{equation}
2\frac{\alpha_s C_F }{\pi^2}  \int^{P_\perp^2}_{q_\perp^2} \frac{d^2l_\perp}{l_\perp^2} \int^{P_\perp}_{l_\perp} \frac{dl^+}{l^+}=\frac{\alpha_s C_F }{2\pi} \ln^2\frac{P_\perp^2}{q_\perp^2}.
\end{equation}
According to its probabilistic interpretation\cite{Mueller:2013wwa}, the Sudakov factor is just the above result with a minus sign. Here we have used the fact that the contributions from the ultra-violet region cancel among all real diagrams. 

As to the case of final state gluon radiation as shown in Fig.~\ref{ref1}, using the same argument, we find no Sudakov double logarithms as long as the angular deviation $\phi\sim \frac{q_\perp}{P_\perp}$ is less than the jet cone size, which is usually chosen to be of order $1$ in high energy experimental analysis. More explicitly, we find the full double logarithmic contribution 
from final state emissions off the $x_1p_{1}^+$-line to be
\begin{equation}
-\frac{\alpha_s C_F}{\pi} \int_{q_\perp^2}^{P_\perp^2} \frac{d k_\perp^2}{k_\perp^2} \int _{k_\perp}^{\textrm{min}[\sqrt{2} x_1 p_1^+, \sqrt{2} x_1 p_1^+ \frac{k_\perp}{\delta P_\perp }]} \frac{d l^+}{l^+}, \label{fradiation}
\end{equation}
where $k_\perp =l_\perp +\xi_g k_{1\perp}$. It is clear that if $\delta > \phi$ there are no $\ln^2 \frac{P_\perp^2}{q_\perp^2}$ terms in Eq.~(\ref{fradiation}), and when $\phi$ is of order $1$ there are no double logarithms of any variety. However, if one were to take $\delta <\frac{q_\perp}{P_\perp}$, Eq.~(\ref{fradiation}) gives Sudakov $\ln^2 \frac{P_\perp^2}{q_\perp^2}$ terms 
and the $\delta$-dependence disappears.

\section{Factorization Results and Matching Between BFKL and Sudakov Resummations}

\begin{figure}[tbp]
\begin{center}
\includegraphics[height=7cm]{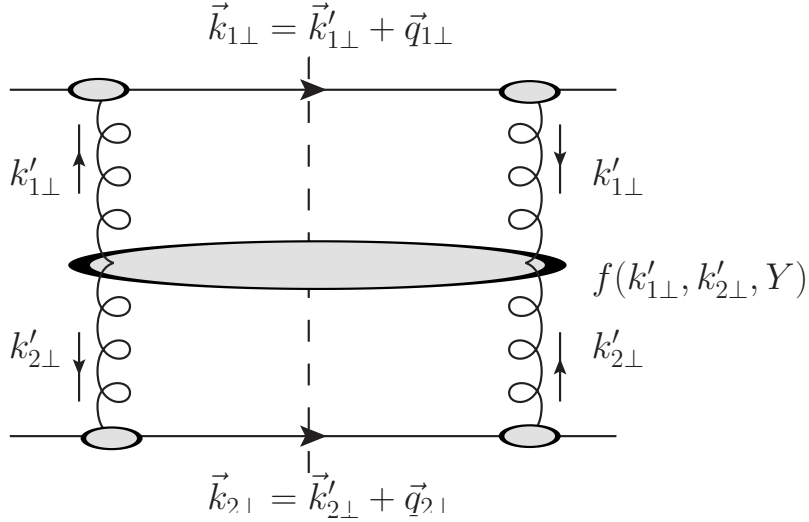}
\end{center}
\caption[*]{Illustration of the factorization formalism for the Mueller-Navelet dijet production:
$q_{1\perp}$ and $q_{2\perp}$ are generated by Sudakov effects and can be related
to the transverse momentum distributions from the incoming nucleons; $f(k_{1\perp}',k_{2\perp}';Y)$
obeys the BFKL evolution, and the associated resummation will be important for large rapidity 
separation of the two jets $Y$.}
\label{fac}
\end{figure}

The Sudakov double logarithms derived in previous sections can be casted into
a factorization formalism. Generic arguments are as follows: incoming partons
contribute to a finite transverse momentum $\vec{q}_\perp$ from collinear
and soft gluon radiations. These radiations are controlled by the Sudakov
formalism, and can be derived formally by the Collins-Soper-Sterman resummation~\cite{Collins:1984kg}.
Each of the incoming partons acquires a final transverse momentum $q_{i\perp}$, 
before they they scatter off each other by exchanging a $t$-channel gluon. The latter
process is dominated by the BFKL dynamics and we can resum the large logarithms
by solving the BFKL evolution equation. Schematically, this can be illustrated 
as Fig.~\ref{fac}.

According to this factorization argument, we can write the differential cross section
for the MN-dijet production as,
\begin{eqnarray}
\frac{d\sigma}{dy_1dy_2d^2k_{1\perp}d^2k_{2\perp}}&=&\int d^2q_{1\perp}d^2q_{2\perp}{\cal F}_a(x_1,q_{1\perp};\mu=k_{1\perp})
{\cal F}_b(x_2,q_{2\perp};\mu=k_{2\perp})\nonumber\\
&&\times \hat\sigma_{ab}(k_{1\perp},k_{2\perp};\mu)f_{BFKL}(\vec{k}_{1\perp}-\vec{q}_{1\perp},\vec{k}_{2\perp}-\vec{q}_{2\perp};Y) \ ,
\end{eqnarray}
where $\hat\sigma_{ab}$ represents the partonic cross section for the $ab$ channel normalized 
with the appropriate color factor in $\sigma_0$ in Eq.~(2). In the above equation,
${\cal F}_{a,b}$ are the so-called transverse momentum distributions (TMDs)
with Sudakov resummation effects including initial and final state radiations~\cite{Collins}.
There is scheme dependence in the TMDs, which, however, will be cancelled by the
associated hard coefficients ${\cal H}$. In the final factorization formula, we choose the
TMDs calculated in the ``TMD"-scheme~\cite{Collins,Prokudin:2015ysa} (or ``Hard" scheme in Ref.~\cite{Catani:2000vq}).
In the dijet production process, final state radiation will also contribute
to the single logarithms which will depend on the jet cone size. 
In general kinematics of dijet production, the TMD resummation is much
more complicated than the above equation, where a matrix form has to 
be included to take into account final state radiation contributions~\cite{Sun:2014gfa,Sun:2015doa}. However,
in the current case, because of the two jets are produced with large rapidity
separation, the resummation formula will be much simplified. In particular, 
the kinematic variables in the partonic processes have the following 
approximations: $s\sim (-u)\gg (-t)$. The scattering process
is dominated by the diagrams shown in Fig.~1. The detailed 
discussion of this aspect will be presented in the Appendix. Here
we list the final results with respect to the above factorization formula. 

From the results discussed in the Appendix, we modify the TMDs
studied in Drell-Yan processes~\cite{Collins,Prokudin:2015ysa,Catani:2000vq}) for 
the dijet resummation, and now they take the following form,
\begin{eqnarray}
{\cal F}_{q}(x,q_\perp;\mu_F=k_\perp)&=& x\int\frac{d^2R_\perp}{(2\pi)^2}e^{iq_\perp\cdot R_\perp}
	e^{-{\cal S}_{sud}^q(\mu_F=k_\perp,R_\perp)}C\otimes f_{q}(x,\mu_b)\ ,\\
{\cal F}_g(x,q_\perp;\mu_F=k_\perp)&=&x\int\frac{d^2R_\perp}{(2\pi)^2}e^{iq_\perp\cdot R_\perp}
e^{-{\cal S}_{sud}^g(\mu_F=k_\perp,R_\perp)}C\otimes f_{g}(x,\mu_b)
\end{eqnarray}
where $f_{q,g}(x,\mu_b)$ are integrated quark/gluon distribution functions at the
scale $\mu_b=c_0/R_\perp$, respectively. In the above equation $C\otimes f_{q,g}$
represent the convolution integral for the parton distributions,
\begin{eqnarray}
C\otimes f_q(x,\mu)&=&\int\frac{dx'}{x'}\sum_iC_{q/i}(x/x')f_i(x',\mu)\ , \\
C\otimes f_g(x,\mu)&=&\int\frac{dx'}{x'}\sum_iC_{g/i}(x/x')f_i(x',\mu)\ , 
\end{eqnarray}
where $i$ runs through all parton flavors including quarks and gluons.
The Sudakov form factor ${\cal S}_{sud}$ contains the final jet 
contributions depending on the jet size $R$ as well. They read as
\begin{equation}
{\cal S}_{sud}^a=\int_{c_0^2/R_\perp^2}^{k_\perp^2}\frac{d\mu^2}{\mu^2}
	\left(A_a\ln\frac{k_{\perp}^2}{\mu^2}+B_a+D_a\ln\frac{1}{R^2}\right) \ ,
\end{equation}
for quark and gluon, respectively. At one-loop order, we have 
\begin{eqnarray}
&&A_q=\frac{\alpha_s}{2\pi}C_F\ ,~~~ A_g=\frac{\alpha_s}{2\pi}C_A\ , \\
&&B_q=-\frac{\alpha_s}{2\pi}\frac{3}{2}C_F  \ , ~~~B_g=-\frac{\alpha_s}{2\pi}2N_c\beta_0 \ ,\\
&&C_{q/q}(x)= 1+\frac{\alpha_s}{2\pi}(1-x)  \ , ~~~C_{g/g}(x)=1+{\cal O}(\alpha_s^2) \ ,\\
&&C_{g/q}(x)=\frac{\alpha_s}{2\pi}x  \ , ~~~C_{q/g}(x)=\frac{\alpha_s}{2\pi}x(1-x) \ ,\\
&&D_q=\frac{\alpha_s}{2\pi}C_F\ , ~~~D_g=\frac{\alpha_s}{2\pi}C_A\ ,
\end{eqnarray}
where $\beta_0=\frac{11}{12}-\frac{N_f}{6N_c}$. The double logarithms presented here are identical to 
those derived in previous sections obtained from different methods. In the phenomenological calculations, we will also introduce the non-perturbative part, see, for example, with $b_*$-prescription~\cite{Collins:1984kg}.
Finally, we shall have one-loop corrections for the hard coefficients,
\begin{eqnarray}
\hat\sigma_{qq'\to qq'}&=&\hat\sigma_{qq'\to qq'}^{(0)}\left\{1+\frac{\alpha_s}{2\pi}\left[2{\cal K}+2\Delta I_q\right] \right\}\ ,\\
\hat\sigma_{qg\to qg}&=&\hat\sigma_{qg\to qg}^{(0)}\left\{1+\frac{\alpha_s}{2\pi}\left[2{\cal K}+\Delta I_q+\Delta I_g\right] \right\}\ ,\\
\hat\sigma_{gg\to gg}&=&\hat\sigma_{gg\to gg}^{(0)}\left\{1+\frac{\alpha_s}{2\pi}\left[2{\cal K}+2\Delta I_g\right] \right\}\ ,
\end{eqnarray}
where ${\cal K}$ and $\Delta I_{q,g}$ are defined as
\begin{eqnarray}
{\cal K}&=&C_A\left(\frac{67}{18}-\frac{\pi^2}{6}\right)-\frac{5N_f}{9}\ ,\\
\Delta I_q&=&C_F\left[\frac{3}{2}\ln\frac{1}{R^2}+\frac{3}{4}+\frac{2}{3}\pi^2\right]\ ,\\
\Delta I_g&=&C_A\left(2\beta_0\ln\frac{1}{R^2}-\frac{\pi^2}{6}\right)-\frac{N_f}{6} \ .
\end{eqnarray} 
The detailed derivation of the above coefficients is presented in the Appendix. It is interesting to
notice that all the partonic channels contain the same correction terms ${\cal K}$ as
what was found in Ref.~\cite{Ciafaloni:1998hu}. In our calculations, we 
have taken into account the anti-$k_t$ algorithm to define the final state jets, where
extra terms are found in association of the final state quark or gluon jets. By 
doing that, we also introduce the jet size-dependent terms in both the Sudakov
form factors and the hard coefficients. However, these are universal, in the sense
that the quark final state contributes to the same factor in both $qg$ and $qq'$ channels.

We can also write down the above expressions in the $R_\perp$-space,
\begin{eqnarray}
\frac{d\sigma}{dy_1dy_2d^2k_{1\perp}d^2k_{2\perp}}&=&\int \frac{d^2\rho_{1\perp}d^2\rho_{2\perp}}{(2\pi)^4}
e^{ik_{1\perp}\cdot \rho_{1\perp}+ik_{2\perp}\cdot \rho_{2\perp}}C\otimes f_{a}(x_1,\rho_{1\perp})C\otimes f_{b}(x_2,\rho_{2\perp})
\nonumber\\
&&\times e^{-{\cal S}_{sud}^a(k_{1\perp},\rho_{1\perp})}e^{-{\cal S}_{sud}^b(k_{2\perp},\rho_{2\perp})}
{\cal H}_{ab}(\mu)T(\rho_{1\perp},\rho_{2\perp};Y) \ ,
\end{eqnarray}
where we have both Sudakov and BFKL resummation effects. Again, ${\cal H}_{ab}$ are hard 
coefficients normalized to the leading order expressions in Sec.~II. From the above results, we find
the one-loop corrections as,
\begin{eqnarray}
{\cal H}_{qq'\to qq'}&=&{\cal H}_{qq'\to qq'}^{(0)}\left\{1+\frac{\alpha_s}{2\pi}\left[2{\cal K}+2\Delta I_q\right] \right\}\ ,\\
{\cal H}_{qg\to qg}&=&{\cal H}_{qg\to qg}^{(0)}\left\{1+\frac{\alpha_s}{2\pi}\left[2{\cal K}+\Delta I_q+\Delta I_g\right] \right\}\ ,\\
{\cal H}_{gg\to gg}&=&{\cal H}_{gg\to gg}^{(0)}\left\{1+\frac{\alpha_s}{2\pi}\left[2{\cal K}+2\Delta I_g\right] \right\}\ ,
\end{eqnarray}
where ${\cal K}$ and $\Delta I_{q,g}$ are defined above.
On the other hand, when the rapidity interval $Y$ is small, we do not need to do BFKL resummation. Therefore, we can replace the last factor in the above equation by 
\begin{equation}
T(\rho_{1\perp},\rho_{2\perp};Y) \Longrightarrow \delta^{(2)}(\rho_{1\perp}-\rho_{2\perp})e^{-Y\int_{c_0/\rho_{1\perp}}^{P_\perp}\frac{d\mu}{\mu}2C_A \frac{\alpha_s}{2\pi}} \ ,
\end{equation}
which reduces to results obtained in the collinear factorization approach. In this case, it seems that the rapidity resummation now can be included in the Sudakov resummation as well. Clearly, the difference between the BFKL and Sudakov resummation relies on the above factor. 

However, the derivation of the Sudakov logarithms is only valid in the region of small $\vec{q}_\perp=\vec{k}_{1\perp}+\vec{k}_{2\perp}$,
where the two jets are produced back-to-back in azimuthal angular distributions. When
the two jets are produced away from back-to-back region, $q_\perp$ is not small compared to $k_{i\perp}$
anymore, and we have to match to the complete BFKL factorization calculations. The latter
has been worked out at the next-to-leading logarithmic order, and the differential
cross section is written as
\begin{eqnarray}
\frac{d\sigma}{dy_1dy_2d^2k_{1\perp}d^2k_{2\perp}}&=&\int \frac{d^2\rho_{1\perp}d^2\rho_{2\perp}}{(2\pi)^4}
e^{ik_{1\perp}\cdot \rho_{1\perp}+ik_{2\perp}\cdot \rho_{2\perp}} x_1f_{a}(x_1,\mu) x_2f_{b}(x_2,\mu)\nonumber\\
&&\times {\cal H}_{ab}^{BFKL}(x_1,x_2,\rho_{1\perp},\rho_{2\perp};\mu)T(\rho_{1\perp},\rho_{2\perp};Y) \ .
\end{eqnarray}
At intermediate $q_\perp$, we expect the above results to match each other.

\section{Discussion and Conclusion}

For dijet production in high energy proton proton collisions, dijets with large rapidity separation are particularly interesting for the study of QCD resummation physics. It has long been realized that the so-called BFKL resummation will be important in this process, which is referred as the Mueller-Navelet dijet
production~\cite{Mueller:1986ey}. On top of the BFKL resummation, through one-loop calculation for this process from different perspectives, we have demonstrated that there should be the resummation of Sudakov factors. In this work, we have obtained various Sudakov double logarithms for Mueller-Navelet jets production in different channels when dijets are almost back-to-back. We believe this results can help to quantitatively study the BFKL dynamics through Mueller-Navelet jets production at the LHC. 

Last but not least, we would like to comment on the recent studies of the transverse momentum resummation for generic dijet production in hardon-hadron collisions\cite{Sun:2014gfa,Sun:2015doa}. It has been shown that the Sudakov resummation is playing an important role in describing the dijet correlation data at both the Tevatron and the LHC. The result presented in this manuscript, which is specifically for Mueller-Navelet dijets with large rapidity separation, is complementary to those studies. 

 \begin{acknowledgments}
This work was supported in part by the U.S. Department of Energy under the contracts DE-AC02-05CH11231. B.X. wishes to thank Dr. F. Yuan and the nuclear theory group at LBNL for hospitality and support during his visit when this work is initiated. L.Sz. was supported by grant of National Science Center, Poland, No. 2015/17/B/ST2/01838.
\end{acknowledgments}

\appendix

\section{Collinear Framework Calculations}

In this appendix, starting from the collinear factorization framework, we would like to
argue that there should be Sudakov resummation as well as the BFKL resummation 
for Mueller-Navelet jets production at high energy colliders. 
We can perform the calculations of dijet production in
the back-to-back correlation region at one-loop order, and
take the high energy limit. From these calculations, we identify the
Sudakov double logarithms, and the BFKL-type logarithms
depending on the rapidity difference $Y$ between the two
jets in the final state. Therefore, we need to perform 
both resummations. 

In Ref.~\cite{Sun:2014gfa,Sun:2015doa}, the Sudakov resummation was derived for dijet production,
which is valid for the two jets produced at the same rapidity region. However,
the derivations can also help us to identify the large logarithms for the Mueller-Navelet
dijet production. In the following, we will extend the calculations in Ref.~\cite{Sun:2014gfa,Sun:2015doa} to
the current case. In addition, 
when the two jets are produced with large rapidity separation, we are
in a special kinematic region, where the physics is dominated by $t$-channel
diagrams. Therefore, we will apply the following kinematic approximations,
$s\sim -u\gg -t $, which also implies that $P_\perp^2\simeq tu/s\approx -t$.
More importantly, all the partonic channels with $t$-channel gluon exchange
will be the most important contributions. This is because they all have terms 
which are proportional to $s^2/t^2$. Therefore, we will only take these dominant 
channels in the following calculations: $qq'\to qq'$, $qg\to qg$, and $gg\to gg$. 
After taking the above limit, the leading order result (Eq.(13) of Ref.~\cite{Sun:2015doa}) 
agrees with Eq.~(2) with leading order expression for 
$f(k_{1\perp},k_{2\perp},Y)|^{LO}=\delta^{(2)}(k_{1\perp}+k_{2\perp})$. Therefore, we do
have the same normalization. In Ref.~\cite{Sun:2015doa}, the differential
cross section for dijet production at the back-to-back correlation limit is calculated 
at one-loop order, taking into account the most important collinear and 
soft gluon radiation contributions. In the collinear calculation set-up, 
$\vec{q}_\perp=\vec{k}_{1\perp}+\vec{k}_{2\perp}$ is the relevant variable
in the final resummation, and one-gluon radiation contributes to non-zero $q_\perp$.

Furthermore, we take the back-to-back correlation limit, i.e., $P_\perp\gg q_\perp$. The 
leading contributions from collinear and soft gluon radiations can be obtained
from the results of Ref.~\cite{Sun:2014gfa,Sun:2015doa}. It becomes much simpler because
of the large rapidity separation of the dijet and the approximation we are 
taking $s\sim -u\gg -t $. For example, for the $qq'\to qq'$ channel, from Eqs.~(65) and (66) of Ref.~\cite{Sun:2015doa},
we have the following expression for the soft gluon contribution at small $q_\perp$ from real diagrams,
\begin{eqnarray}
\frac{\alpha_s}{2\pi^2}\frac{1}{q_{\perp}^2}\left\{2C_F\ln\frac{P_\perp^2}{q_{\perp}^2}+2C_F\ln\frac{1}{R^2}
+2C_A\ln\frac{s}{P_\perp^2}\right\} \ ,
\end{eqnarray}
where the first term corresponds to the Sudakov double logs, the second term for the
jet functions, and the third term for the BFKL small-x resummation term. 
To derive the above results, we have applied the anti-$k_t$ algorithm for the
final state jets. When the gluon radiation is inside the jet cone, it will not contribute
to the finite $q_\perp$. This requirement leads to the jet size dependent term in the
above equation~\cite{Sun:2014gfa,Sun:2015doa}.
We can also identify the third term depending on the rapidity separation 
between the two jets,
$\ln({s}/{P_\perp^2})\sim Y $, where $Y$ is the rapidity difference between
the two jets. Similarly, from the results in Ref.~\cite{Ellis:1985er}, the virtual contribution can be simplified as,
\begin{eqnarray}
&&\frac{\alpha_s}{2\pi}\left\{C_F\left[-\frac{4}{\epsilon^2}+\frac{1}{\epsilon}\left(4\ln\frac{P_\perp^2}{\mu^2}-6\right)\right]
+C_A\frac{1}{\epsilon}2\ln\frac{s}{P_\perp^2}+C_F\left(-2\ln^2\left(\frac{s}{P_\perp^2}\right)-6\ln\frac{s}{P_\perp^2}-16\right)
\nonumber\right.\\
&&\left.~~~+C_A\left(2\ln^2\left(\frac{s}{P_\perp^2}\right)+\pi^2+\frac{85}{9}- 2\beta_0\ln\frac{P_\perp^2}{\mu_R^2}\right)-\frac{20}{9}\frac{N_f}{2}\right\}\ .
\end{eqnarray}
To obtain the complete one-loop result, we Fourier transform $q_\perp$-dependent
expressions to $R_\perp$-space, and add the virtual contribution,
\begin{eqnarray}
\widetilde{W}_{qq'\to qq'}^{(1)}&=&\frac{\alpha_s}{2\pi}\left\{
-\ln\left(\frac{\mu^2R_\perp^2}{c_0^2}\right)\left[{\cal P}_{qq}(\xi)\delta(1-\xi')
+{\cal P}_{qq}(\xi')\delta(1-\xi)\right] +\delta(1-\xi)\delta(1-\xi')\right.\nonumber\\
&&\times\left[-C_F\ln^2\left(\frac{P_\perp^2R_\perp^2}{c_0^2}\right)-\ln\frac{P_\perp^2R_\perp^2}{c_0^2}\left(C_F
\left(-3+2\ln\frac{1}{R^2}\right)+2C_A\ln\frac{s}{P_\perp^2}\right)\right.\nonumber\\
&&\left.\left.+C_A\left(\frac{85}{9}+\pi^2\right)+C_F\left(3\ln\frac{1}{R^2}-3-\frac{5\pi^2}{3}\right)-\frac{20}{9}\frac{N_f}{2}
\right]\right\} \ ,
\end{eqnarray}
where we have included the collinear gluon contributions associated with the
two incoming quark distributions. We have also set the renormalization scale
for the running coupling constant at $P_\perp$ to simplify the above 
expression, $\mu_R=P_\perp$. $\widetilde{W}(R_\perp)$ corresponds to 
the Fourier transform of $f(k_{1\perp},k_{2\perp},Y)$ in Eq.~(2) from the 
collinear factorization calculations. 
Again, we can clearly identify the three important terms in the 
above equation: Sudakov double logarithms, single logarithms associated
with collinear gluon radiation contribution, and the BFKL-term.

Following the Sudakov resummation procedure~\cite{Sun:2014gfa,Sun:2015doa}, we would arrive at the following
resummation result,
\begin{eqnarray}
W_{qq'\to qq'}(R_\perp)&=&x_1f_q(x_1,c_0/R_\perp)x_2f_{q'}(x_2,c_0/R_\perp)h_{qq'\to qq'}
e^{-\widetilde{S}_{qq'\to qq'}(s,P_\perp,R_\perp)} \ ,
\end{eqnarray}
where the simplified Sudakov form factor is defined as
\begin{equation}
\widetilde{S}_{qq'\to qq'}=\int^{P_\perp^2}_{c_0^2/R_\perp^2}\frac{d\mu^2}{\mu^2}\frac{\alpha_s(\mu)}{2\pi}
\left[2C_F\ln\left(\frac{P_\perp^2}{\mu^2}\right)-3C_F+2C_F\ln\frac{1}{R^2}+2C_A\ln\frac{s}{P_\perp^2}\right]\ . \label{sufqq1}
\end{equation}
By using the above results, we will find that the hard coefficient can be
written as
\begin{equation}
h_{qq'\to qq'}=h_{qq'\to qq'}^{(0)}\left\{1+\frac{\alpha_s}{2\pi}\left[C_A\left(\frac{85}{9}
+\pi^2\right)+C_F\left(3\ln\frac{1}{R^2}-3-\frac{5\pi^2}{3}\right)-\frac{20}{9}\frac{N_f}{2}\right]\right\} \ .
\end{equation}
It is interesting to note that the above equation can also be written as
follows,
\begin{equation}
h_{qq'\to qq'}=h_{qq'\to qq'}^{(0)}\left\{1+\frac{\alpha_s}{2\pi}\left[2{\cal K}+2\Delta I_q\right]\right\}\ ,
\end{equation}
where the same coefficient 
\begin{equation}
{\cal K}=C_A\left(\frac{67}{18}-\frac{\pi^2}{6}\right)-\frac{5N_f}{9}\ ,
\end{equation} 
appears in Ref.~\cite{Ciafaloni:1998hu}, and the extra terms come
from the different treatment of the final state jets,
\begin{equation}
\Delta I_q=C_F\left[\frac{3}{2}\ln\frac{1}{R^2}+\frac{3}{4}+\frac{2}{3}\pi^2\right]\ .
\end{equation}
In our calculations, as mentioned above, we take the anti-$k_t$ algorithm to derive
the final state jet contributions, whereas the whole phase space was integrated out
in Ref.~\cite{Ciafaloni:1998hu}. We note that different jet algorithm will lead to
different results in the above equation.

The calculations for $qg\to qg$  and $gg\to gg$ channels 
can be followed accordingly. 
For the $qg\to qg$ channel, we have the real diagram contribution,
\begin{eqnarray}
\frac{\alpha_s}{2\pi^2}\frac{1}{q_{\perp}^2}\left\{\left(C_A+C_F\right)\ln\frac{P_\perp^2}{q_{\perp}^2}+\left[C_A\ln\frac{1}{R^2}+C_F\ln\frac{1}{R^2}\right]
+2C_A\ln\frac{s}{P_\perp^2}\right\}\ ,
\end{eqnarray}
and the virtual contribution reads~\cite{Ellis:1985er},
\begin{eqnarray}
&&\frac{\alpha_s}{2\pi}\left\{-\frac{2}{\epsilon^2}(C_A+C_F)
+\frac{2}{\epsilon}\left(C_A\ln\frac{s}{\mu^2}-\frac{3}{2}C_F-2\beta_0C_A+C_F\ln\frac{P_\perp^2}{\mu^2}\right)\right.\nonumber\\
&&+(C_A+C_F)\left(-\ln^2\left(\frac{s}{\mu^2}\right)+\frac{\pi^2}{6}\right)
+C_F\left(2\ln\frac{s}{\mu^2}\ln\frac{s}{P_\perp^2}+3\ln\frac{P_\perp^2}{\mu^2}-8\right)\nonumber\\
&&\left.+(C_A-C_F)\ln^2\left(\frac{s}{P_\perp^2}\right)+C_A(\pi^2+1)\right\} \ .
\end{eqnarray}
By adding up soft and jet contributions, we obtain the full one-loop result
for $W(b)$ as
\begin{eqnarray}
\widetilde{W}_{qg\to qg}^{(1)}&=&\frac{\alpha_s}{2\pi}\left\{-\ln\left(\frac{\mu^2R_\perp^2}{c_0^2}\right)\left[{\cal P}_{gg}(\xi)\delta(1-\xi')
+{\cal P}_{gg}(\xi')\delta(1-\xi)\right] +\delta(1-\xi)\delta(1-\xi')\right.\nonumber\\
&&\times \left[C_F\left(-\frac{1}{2}\ln^2\left(\frac{P_\perp^2R_\perp^2}{c_0^2}\right)-\ln\frac{P_\perp^2R_\perp^2}{c_0^2}
\left(\ln\frac{1}{R^2}-\frac{3}{2}\right)+\frac{3}{2}\ln\frac{1}{R^2}-\frac{3}{2}-\frac{5\pi^2}{6}\right)\right.\nonumber\\
&&+C_A\left(-\frac{1}{2}\ln^2\left(\frac{P_\perp^2R_\perp^2}{c_0^2}\right)-\ln\frac{P_\perp^2R_\perp^2}{c_0^2}
\left(\ln\frac{1}{R^2}-2\beta_0+2\ln\frac{s}{P_\perp^2}\right)\right.\nonumber\\
&&\left.\left.\left.+2\beta_0\ln\frac{1}{R^2}+\frac{\pi^2}{6}+\frac{76}{9}-\frac{23N_f}{54}\right)\right]\right\} \ .
\end{eqnarray}
From the above results, we derive the Sudakov resummation formula,
\begin{eqnarray}
W_{qg\to qg}(R_\perp)&=&x_1f_q(x_1,c_0/R_\perp)x_2f_{g}(x_2,c_0/R_\perp)h_{qg\to qg}
e^{-\widetilde{S}_{qg\to qg}(s,P_\perp,R_\perp)} \ ,
\end{eqnarray}
where the simplified Sudakov form factor is defined as
\begin{eqnarray}
\widetilde{S}_{qg\to qg}&=&\int^{P_\perp^2}_{c_0^2/R_\perp^2}\frac{d\mu^2}{\mu^2}\frac{\alpha_s(\mu)}{2\pi}
\left[\ln\left(\frac{P_\perp^2}{\mu^2}\right)(C_A+C_F)-\frac{3}{2}C_F-2\beta_0C_A\right.\nonumber\\
&&\left.~~~~+C_F\ln\frac{1}{R^2}+C_A\ln\frac{1}{R^2}+2C_A\ln\frac{s}{P_\perp^2}\right]\ , \label{sufqg}
\end{eqnarray}
and the hard coefficient is calculated as
\begin{eqnarray}
h_{qg\to qg}&=&h_{qg\to qg}^{(0)}\left\{1+\frac{\alpha_s}{2\pi}\left[C_F\left(\frac{3}{2}\ln\frac{1}{R^2}-\frac{3}{2}-\frac{5\pi^2}{6}\right)
\right.\right.\nonumber\\
&&\left.\left.~~~~+C_A\left(2\beta_0\ln\frac{1}{R^2}+\frac{\pi^2}{6}+\frac{76}{9}-\frac{23N_f}{54}\right)\right]\right\} \ .
\end{eqnarray}
Again, if we write in terms of ${\cal K}$, we have the following result for $h_{qg\to qg}$,
\begin{eqnarray}
h_{qg\to qg}=h_{qg\to qg}^{(0)}\left\{1+\frac{\alpha_s}{2\pi}\left[2{\cal K}+\Delta I_q+\Delta I_g\right] \right\}\ .
\end{eqnarray}
Because we have a quark jet plus a gluon jet in the final state, the extra terms differ from the above
$qq'\to qq'$ channel, and the gluon term reads as
\begin{equation}
\Delta I_g=C_A\left(2\beta_0\ln\frac{1}{R^2}-\frac{\pi^2}{6}\right)-\frac{N_f}{6} \ ,
\end{equation}
which comes from the gluon jet contribution.

For $gg\to gg$ channel, we have the real diagram contribution,
\begin{eqnarray}
\frac{\alpha_s}{2\pi^2}\frac{1}{q_{\perp}^2}\left\{2C_A\ln\frac{P_\perp^2}{q_{\perp}^2}+2C_A\ln\frac{1}{R^2}
+2C_A\ln\frac{s}{P_\perp^2}\right\}\ .
\end{eqnarray}
The virtual graph contribution is simplified as~\cite{Ellis:1985er}
\begin{eqnarray}
&&\frac{\alpha_s}{2\pi}C_A\left\{-\frac{4}{\epsilon^2}+\frac{1}{\epsilon}\left(2\ln\frac{s}{\mu^2}
+2\ln\frac{P_\perp^2}{\mu^2}-8\beta_0\right)+8\beta_0\ln\frac{P_\perp^2}{\mu^2}\right.\nonumber\\
&&~~~\left.-4\beta_0\ln\frac{P_\perp^2}{\mu_R^2}
-2\ln\frac{s}{\mu^2}\ln\frac{P_\perp^2}{\mu^2}+\frac{10N_f}{27}+\frac{4}{3}\pi^2-\frac{67}{9}\right\} \ .
\end{eqnarray}
Adding the soft and jet contributions, we obtain the total contribution for $W(b)$
at one-loop order,
\begin{eqnarray}
\widetilde{W}_{gg\to gg}^{(1)}&=&\frac{\alpha_s}{2\pi}\left\{-\ln\left(\frac{\mu^2R_\perp^2}{c_0^2}\right)\left[{\cal P}_{gg}(\xi)\delta(1-\xi')
+{\cal P}_{gg}(\xi')\delta(1-\xi)\right] +\delta(1-\xi)\delta(1-\xi')\right.\nonumber\\
&&\times C_A\left[ -\ln^2\left(\frac{P_\perp^2R_\perp^2}{c_0^2}\right)-\ln\frac{P_\perp^2R_\perp^2}{c_0^2}
\left(2\ln\frac{s}{P_\perp^2}+2\ln\frac{1}{R^2}\right)\right.\nonumber\\
&&\left.\left.+2\beta_0\left(2\ln\frac{P_\perp^2R_\perp^2}{c_0^2}
+2\ln\frac{1}{R^2}\right)
+\frac{67}{9}-\frac{13N_f}{27}-\frac{2\pi^2}{3}\right]\right\} \ .
\end{eqnarray}
From the above results, we obtain the Sudakov resummation as
\begin{eqnarray}
W_{gg\to gg}(R_\perp)&=&x_1f_q(x_1,c_0/R_\perp)x_2f_{g}(x_2,c_0/R_\perp)h_{gg\to gg}
e^{-\widetilde{S}_{gg\to gg}(s,P_\perp,R_\perp)} \ .
\end{eqnarray}
where the simplified Sudakov form factor is defined as
\begin{equation}
\widetilde{S}_{gg\to gg}=\int^{P_\perp^2}_{c_0^2/R_\perp^2}\frac{d\mu^2}{\mu^2}\frac{\alpha_s(\mu)}{2\pi}
\left[\ln\left(\frac{P_\perp^2}{\mu^2}\right)2C_A-4\beta_0C_A+2C_A\ln\frac{1}{R^2}+2C_A\ln\frac{s}{P_\perp^2}\right]\ , \label{sufgg}
\end{equation}
with hard coefficient as
\begin{equation}
h_{gg\to gg}=h_{gg\to gg}^{(0)}\left\{1+\frac{\alpha_s}{2\pi}C_A
\left[4\beta_0\ln\frac{1}{R^2}+\frac{67}{9}-\frac{13N_f}{27}-\frac{2\pi^2}{3}\right]\right\} \ .
\end{equation}
It can also be written as
\begin{equation}
h_{gg\to gg}=h_{gg\to gg}^{(0)}\left\{1+\frac{\alpha_s}{2\pi}\left[2{\cal K}+2\Delta I_g\right]\right\} \ ,
\end{equation}
where $\Delta I_g$ is defined above.

We would like to emphasize that the above resummation results show that
large logarithms $\ln (s/P_\perp^2)$ play an important role for dijet production
with large rapidity separation. We need to separately resum these large
logarithms. In addition, because of the t-channel gluon exchange dominance, this term
is universal among different channels. 
This can be seen from Eqs.~(\ref{sufqq1},\ref{sufqg},\ref{sufgg}).
It is also consistent with a factorization in terms of BFKL
resummation.

\end{document}